\documentclass[12pt,preprint]{aastex}

\newcommand{\pms}{pre--main-sequence}
\newcommand{\ha}{H$\alpha$}
\newcommand{\kms}{km s$^{-1}$}
\newcommand{\martin}{Mart\'{\i}n}
\shorttitle{Pulsed accretion in UZ Tau E}
\shortauthors{Jensen et al.}
\slugcomment{To appear in the Astronomical Journal, July 2007}

\defcitealias{AL96}{AL96}

\begin{document}

\title{Periodic accretion from a circumbinary disk in the young binary
  UZ Tau E}

\author{Eric L. N. Jensen\altaffilmark{1}, Saurav 
  Dhital\altaffilmark{1,2}, Keivan G. Stassun\altaffilmark{2},
  Jenny Patience\altaffilmark{3}, 
  William Herbst\altaffilmark{4}, 
  Frederick M. Walter\altaffilmark{5}, 
  Michal Simon\altaffilmark{5},
  Gibor Basri\altaffilmark{6}
}

\altaffiltext{1}{Swarthmore College Department of Physics and
  Astronomy, 500 College Ave., Swarthmore PA 19081;
  ejensen1@swarthmore.edu.}

\altaffiltext{2}{Department of Physics and Astronomy, Vanderbilt
  University, Nashville, TN 37235}
\altaffiltext{3}{Division of Physics, Mathematics, and Astronomy,
  California Institute of Technology, Pasadena, CA 91125}
\altaffiltext{4}{Astronomy Department, Wesleyan University,
  Middletown, CT 06459}
\altaffiltext{5}{Department of Physics and Astronomy, State University
of New York, Stony Brook, NY 11794-3800}
\altaffiltext{6}{Astronomy Department, University of California, 521
  Campbell Hall, Berkeley, CA 94720}

\begin{abstract}
  Close \pms\ binary stars are expected to clear central holes in
  their protoplanetary disks, but the extent to which material can
  flow from the circumbinary disk across the gap onto the individual
  circumstellar disks has been unclear.  In binaries with eccentric
  orbits, periodic perturbation of the outer disk is predicted to
  induce mass flow across the gap, resulting in accretion that varies
  with the binary period.  This accretion may manifest itself
  observationally as periodic changes in luminosity.  Here we present
  a search for such periodic accretion in the \pms\ spectroscopic
  binary UZ Tau E\null.  We present $BVRI$ photometry spanning three
  years; we find that the brightness of UZ Tau E is clearly periodic,
  with a best-fit period of $19.16 \pm 0.04$ days.  This is consistent
  with the spectroscopic binary period of $19.13$ days, refined here
  from analysis of new and existing radial velocity data.  The
  brightness of UZ Tau E shows significant random variability, but the
  overall periodic pattern is a broad peak in enhanced brightness,
  spanning more than half the binary orbital period.  The variability
  of the H$\alpha$ line is not as clearly periodic, but given the
  sparseness of the data, some periodic component is not ruled out.
  The photometric variations are in good agreement with predictions
  from simulations of binaries with orbital parameters similar to
  those of UZ Tau E, suggesting that periodic accretion does occur
  from circumbinary disks, replenishing the inner circumstellar disks
  and possibly extending the timescale over which they might form
  planets.
\end{abstract}


\section{Introduction}

It is now well-established that most stars are members of binary
systems at birth, and that many of these stars are surrounded by disks
similar to those found around young single stars (see, e.g., the
recent review by \citealt{MoninPPV}).  Thus, understanding the origin
of binaries is vital to understanding the star formation process.  The
predominance of binaries also means that, based on number of systems
alone, most potential sites of planet formation lie in multiple
systems.  However, interactions between stars and disks in binary
systems can alter disk structure
\citep{BSCG,JMF1994,OsterlohBeckwith1995,JKM1996,JMF1996,JensenMathieu1997},
resulting in a more complicated environment for planet formation.
Nonetheless, the discovery of planets in relatively close binary
systems \citep{Eggenberger2004} shows that binary systems are
viable sites of planet formation.  Understanding the extent to which
binaries modify the structure of their surrounding disks is important
for understanding the possible diversity of planetary system
environments.  In addition, though the mass ratios are different, the
interactions between stellar binary companions and disks involve the
same physics as those between planetary companions and disks
\citep[e.g.,][]{DAngelo2006} but are more easily observable.  Thus, an
understanding of binary-disk interactions may help us understand
planet formation around single stars as well.

A binary star system may have up to three disks: two circumstellar
disks, one each around the primary and secondary; and a circumbinary
disk outside the binary orbit.  Both analytic calculations and
numerical simulations show that the region between these disks is not
stable for orbiting disk material.  However, the question of how
easily material can flow from the outer, circumbinary disk across the
gap to the circumstellar disks has not been clear, either from an
observational or a theoretical standpoint.  The spectral energy
distributions of some young binaries show the clear signature of a
cleared central region, while other, apparently similar systems do not
\citep{JensenMathieu1997}. Theoretical analyses by
\cite{1993prpl.conf..749L} and \cite{1994ApJ...421..651A} suggested
that material in the region around the binary orbit is cleared,
creating a quasi-equilibrium structure with three distinct disks and a
cleared region between them.  If the gap between disks is impermeable,
the disks evolve independently of each other.  Since the presence of a
binary companion may increase the rate of accretion
\citep{1992ASPC...32..176C, 1992ApJ...399..192O} and the binary orbit
presents a constraint on the size of each circumstellar disk, the
circumstellar disks would be exhausted much more quickly than disks
around single stars.  However, smoothed particle hydrodynamic
simulations by \cite{AL96} (hereafter AL96; see also
\citealt{GuentherKley2002}) predicted that material may indeed flow
from the circumbinary disk to the circumstellar environment, with the
accretion rate varying with the phase of the binary orbit.

If such periodic accretion occurs in young binaries, it may be
detectable observationally by a periodic brightening of the system as
the material flowing from the circumbinary disk shocks when it
collides with the circumstellar disk(s) or accretes onto the stellar
surface(s).  Observations of the T Tauri spectroscopic binary DQ Tau
by \cite{Mathieu1997} showed such brightening, occurring at the binary
orbital period.  In addition, DQ Tau shows periodic variations in
spectral veiling and emission line intensities with orbital phase
\citep{Basri1997}, providing strong support for the broad picture of
mass flow across gaps suggested by AL96.  However, subsequent searches
in other young, short-period binary systems surrounded by disks have
yielded mixed results.  \citet{2003AA...409.1037A} did not find
periodic photometric variations in AK Sco, but they did find that the
blue wing of the H$\alpha$ line, and both the blue and red wings of
the H$\beta$ line, vary with the binary orbital period, as does the
total H$\alpha$ equivalent width.  V4046 Sgr has shown periodic
photometric variations at the binary orbital period \citep{Quast2000,
  Mekkaden2000}, though an earlier study did not find such variations
\citep{Byrne1986}.  Like AK Sco, however, V4046 Sgr does show
variations in the equivalent width and shape of Balmer lines as a
function of orbital phase \citep{StempelsGahm2004}.  No dedicated
photometric monitoring of UZ Tau E has been reported in the literature
to date, but in previous spectroscopic observations, neither H$\alpha$
equivalent width nor spectral veiling has shown any obvious dependence
on binary orbital phase \citep{Martin2005}.

Motivated by previous observational work and a desire to understand
accretion in binary systems, we have undertaken a photometric
monitoring campaign for the \pms\ (PMS) spectroscopic binary UZ Tau
E\null.  UZ Tau, in the Taurus-Auriga star-forming region, was first
discovered to be variable by Bohlin during a bright outburst in 1921
\citep{Bailey1921,Bohlin1923} and identified as a $\sim 3\farcs7$
binary by \cite{1944PASP...56..123J}, one of the first \pms\ binaries
to be identified.  Subsequently, both components of the binary were
found to be binary themselves, making this a quadruple system.
\cite{1992ApJ...384..212S} and \cite{1993AJ....106.2005G} identified
UZ Tau W as a binary system, and \cite{Mathieu1996} identified UZ Tau
E to be a single-lined spectroscopic binary with a 19.1-day period.
UZ Tau W, a 0\farcs34 binary (47.6 AU assuming a distance of 140 pc to
Taurus-Auriga; \citealt{Kenyon1994}), is separated from UZ Tau E by
3\farcs78 (530 AU; \citealt{1995ApJ...443..625S}).  \citet{Prato2002}
detected absorption lines of the secondary star in the near infrared
spectrum of UZ Tau E, measuring the mass ratio $M_2/M_1=0.28 \pm
0.01$. \cite{Martin2005} presented additional radial velocity data for
UZ Tau E; they found a binary orbital period of 18.979 days and an
eccentricity of 0.14.  UZ Tau E shows strong \ha\ emission, indicative
of ongoing accretion, and strong infrared and millimeter excess
emission from circumstellar and circumbinary disks; the circumbinary
disk has been resolved at $\lambda=1.3$ mm \citep{JKM1996} and 2.6 mm
\citep{1996A&A...309..493D} with a size and mass comparable to disks
around single stars, showing that the close spectroscopic companion
has not significantly decreased the disk mass, in contrast to the
50-AU pair in UZ Tau W, where the presence of a companion at a
separation comparable to typical disk sizes has greatly reduced the
presence of circumstellar material.

In this paper, we present our new photometric observations, as well as
a re-determination of the binary orbital parameters from new and
existing radial velocity observations.  We then examine periodicities
in the data, showing that the photometric data vary at the binary
orbital period.  Finally, we interpret the results in the context of
the model of pulsed accretion in binary systems.

\section{Observations}

\subsection{Photometry}

In order to search for periodic photometric variations, we have
obtained new photometry of UZ Tau E\null.  Our photometric
observations were made with the 0.6-m Perkin Telescope at the Van
Vleck Observatory (VVO) at Wesleyan University, and with ANDICAM on
the 1.3-m SMARTS Telescope at CTIO\null.  See Table
\ref{table:observations} for details of the observations.  The data
were reduced using standard techniques.

Since UZ Tau E and UZ Tau W are separated by 3\farcs78, we sought to
minimize contamination of the UZ Tau E photometry by light from UZ Tau
W by rejecting images with FWHM greater than 7 pixels, and by using a
relatively small (3-pixel-radius) photometric aperture.  Light curves
of UZ Tau E and W show no correlation, indicating that the UZ Tau E
photometry is uncontaminated.

We performed differential photometry on UZ Tau E using USNO-B1.0
1158-0057597 as a comparison star.  The UZ Tau field is relatively
sparse, and in some of the SMARTS images, this was the only star that
was sufficiently bright to serve as a comparison star, so we used it
as the sole comparison in all of our photometry.  The star was
verified to be non-variable at the few percent level, showing a
standard deviation of 0.02 magnitudes by comparison with several other
stars of similar brightness using the wider-field VVO images.  The
USNO-B1.0 Catalog \citep{2003AJ....125..984M} gives magnitudes for
this star from the Second Palomar Sky Survey of $B=16.39$, $R=13.75$,
and $I=12.8$.  Although these filters are not identical to those used
in our CCD observations, we adopted these values for the magnitude of
the reference star to set the zero point of our light curves.  In
addition, we adopted $V=14.99$ by noting that the $R-I$ color for the
comparison star suggests a spectral type of $\sim$ M0, and adopting a
corresponding $B-V$ color.  Adopting these values allows us to
determine approximate colors for UZ Tau E, though we caution that the
absolute scaling of both the individual magnitudes and the colors is
uncertain.  The differential photometry and the color changes, which
form the basis of our analysis, are both unaffected by this systematic
uncertainty in the zero point.

\subsection{Spectroscopy}
\label{section:spectroscopy}

We did not acquire new spectra of UZ Tau E solely for this program,
but we did make new measurements of a number of spectra taken during
the course of other programs.  Some of these spectra were kindly
supplied by Marcos Huerta.  These are echelle spectra from McDonald
observatory spanning 14 nights in January 2002, with $R = 46,000$ and
wavelength coverage of 5460--6760 \AA.  The observations and data
reduction are described in detail in \citet{Huerta2005}.  Additional
echelle spectra of UZ Tau E were taken at Keck ($R = 31,000$) with the
setup described in \citet{BasriReiners2006} and at Lick ($R = 48,000$)
with the setup described in \citet{AlencarBasri2000}.

We used the spectra from Huerta to measure radial velocities of UZ Tau
E\null.  Spectra of the weak-lined T Tauri star V819 Tau were used as
a radial velocity standard.  By cross-correlating the UZ Tau E spectra
against the V819 Tau spectra, we measured heliocentric radial
velocities of UZ Tau E, assuming $v_{\rm helio} = 14.4 \pm 1.5$ km
s$^{-1}$ for V819 Tau \citep{Walter1988}.  The resulting velocities
are given in Table \ref{table:spectra}.  Radial velocities were
measured using several different echelle orders with strong absorption
lines; the quoted uncertainties reflect the dispersion in these
different measurements, as well as the uncertainty of V819 Tau's
radial velocity.

In addition, we also measured the equivalent width of the \ha\ line in
both sets of spectra in order to track changes in accretion rate over
time.  Equivalent widths are given in Table \ref{table:spectra}, with
an estimated uncertainty of 10\%.

\section{Binary orbital parameters}
\label{section:orbit}

In order to assess whether or not any periodic photometric variations
detected in UZ Tau E are synchronized with the binary orbit, we need
to have an accurate knowledge of the orbital parameters.  These have
been determined previously by \citet{Mathieu1996}, \citet{Prato2002},
and \citet{Martin2005}, but the number of radial velocity points
available is still relatively small, especially for the secondary,
leaving open the prospect of further improvements to the orbital
parameters.  To that end, we have re-analyzed the spectroscopic orbit
using data published in \citet{Prato2002} and \citet{Martin2005} as
well as our new radial velocity measurements (Section
\ref{section:spectroscopy}).

\begin{figure}
\plotone{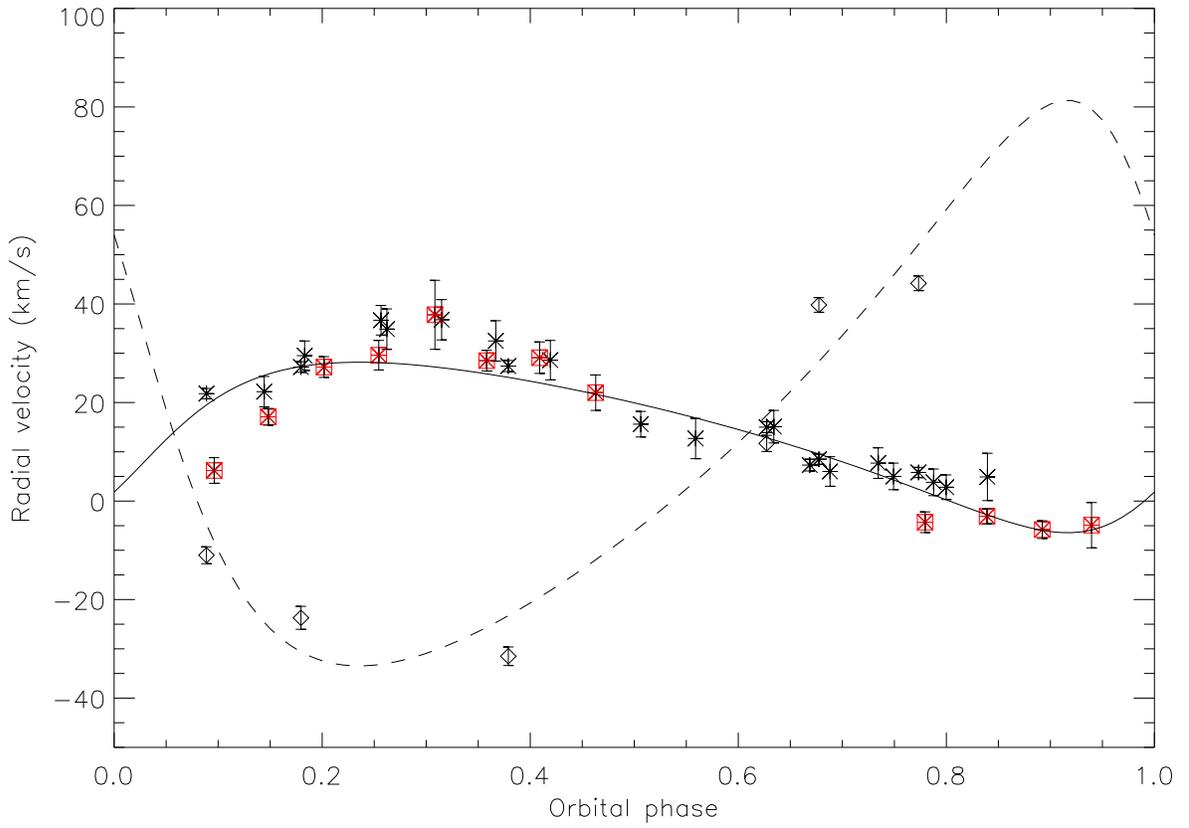}
\caption{The best-fit spectroscopic orbit for UZ Tau E\null.  Crosses
  show velocities of the primary; those enclosed in boxes (in red in
  the on-line edition) show new radial velocity measurements presented
  here.  Open diamonds show velocities of the secondary from \citet{Prato2002}.
  \label{figure:sb-orbit}}
\end{figure}

We fit the radial velocity data using the Binary Star Combined
Solution software \citep{Gudehus2001}, the ORBIT code
\citep{Forveille1999}, and our own custom-written IDL code; all gave
the same solution.  The best-fit phased radial velocity curve is shown
in Figure \ref{figure:sb-orbit} and the orbital parameters are given
in Table \ref{table:orbit}.

The best-fit period of $19.131 \pm 0.003$ days is inconsistent with
the value of $18.979 \pm 0.007$ days found by \citet{Martin2005}.
Examination of the power spectrum of the velocity data used by
\martin\ et al.\ shows that the 18.979-day period appears to be an
alias of the true period, caused by beating of the 19.131-day period
with two six-year gaps in the radial velocity data; there is a
corresponding alias at 19.3 days.  Re-fitting only the data used by
\martin\ et al., we find that the two periods both correspond to local
minima in $\chi^2$ space, with reduced $\chi^2=8.8$ for $P=19.131$
days and reduced $\chi^2=10.7$ for $P=18.979$ days.  When the new
radial velocity data are added, the fit for $P=19.131$ days improves
to reduced $\chi^2=8.1$ while that for $P=18.979$ days worsens to
$\chi^2=11.7$, as expected if 19.131 days is the correct period.

We note that we have not added any additional radial velocity
measurements of the secondary, and thus the mass ratio remains more
uncertain than the other orbital elements, resting on the six
secondary radial velocities presented by \citet{Prato2002}.

UZ Tau E is one of only a handful of pre--main-sequence systems with
measured stellar masses (see \citealt{Mathieu2006} for a recent
review).  Because the total system mass has been measured
\citep{Simon2000}, the spectroscopic orbital parameter $M \sin^3 i$
can be used to determined the orbital inclination.  This can then be
compared with the observed inclination of the circumbinary disk.
While this was done by \citet{Simon2000} and \citet{Prato2002}, we
revisit this issue here using our newly-determined orbital parameters
for UZ Tau E.  Combined with $M = 1.31 \pm 0.08$ $M_\odot$
\citep{Simon2000}, our orbital parameters give $\sin i_{\rm orbit} =
0.81 \pm 0.05$, or $i_{\rm orbit} = 54\arcdeg \pm 5\arcdeg$.  This is
in excellent agreement with the disk inclinations of $54\arcdeg \pm
3\arcdeg$ and $56\arcdeg \pm 2\arcdeg$ measured from interferometric
images of the $\lambda = 1.3$ mm continuum emission and the CO line
emission, respectively \citep{Simon2000}.  Thus, the binary orbit and
the circumbinary disk are coplanar.  Since the disk inclination is
measured at scales of $\sim 100$ AU and the binary orbit is only a few
tenths of an AU, this co-planarity apparently extends over the entire
disk.  Though there are several theoretical studies of how tilted
circumstellar disks interact with a binary system
\citep[e.g.,][]{PapaloizouTerquem1995, Larwood1996, Bate2000,
  LubowOgilvie2000}, we know of no studies of the timescale for
alignment of a circumbinary disk if it is initially tilted relative to
the binary orbit.  Studies of the effects of a planetary-mass companion
on a tilted external disk \citep{LubowOgilvie2001} suggest that such
disk tilts do decay over time, however.  The example of UZ Tau E shows
that co-planarity over the entire disk exists already by an age of a
few Myr, suggesting that circumbinary disks either form already
aligned with the orbit, or come into alignment very quickly.  This is
similar to the result of \citet{Jensen2004}, who found that
circumstellar disks in young binaries tend to be aligned with each
other, and thus presumably with the binary orbit.

\section{Periodic variations}
\label{section:periods}

\subsection{Photometry}

\begin{figure}
\plotone{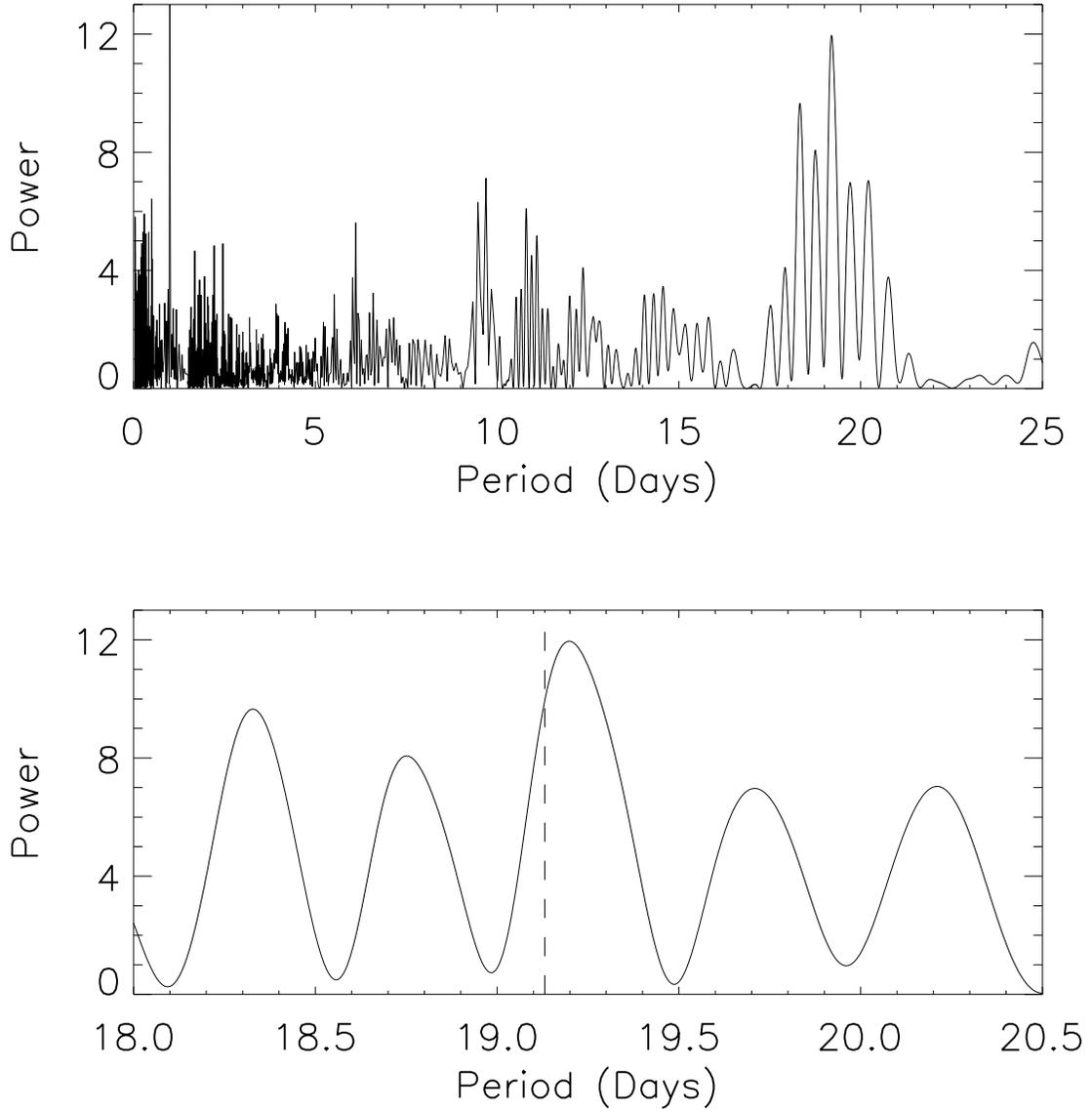}
  \caption{The Lomb-Scargle periodogram for the $I$-band data. The
    periodogram peaks at a period of 19.20 days, with a false-alarm
    probability of 0.001. The dashed line shows the binary orbital
    period of 19.131 days. The smaller peaks visible flanking the main
    peak (lower panel) are near the alias periods expected for beat
    periods between one-year and two-year periods (caused by the
    seasonal gaps in the data) and the binary
    period.\label{figure:periodogram}}
\end{figure}

In order to determine whether the system is varying in phase with the
binary orbit, or in any other systematic way, we have searched the
photometric data for periodic signals.
We begin by searching for evidence of periodicity, without
pre-supposing a particular period.  A Lomb-Scargle periodogram
\citep{Scargle1982} of the $I$-band data (the band with the largest
number of points and best time coverage; Table
\ref{table:observations}) is shown in Figure \ref{figure:periodogram}.
There is a strong peak at a period of $19.20 \pm 0.03$ days.  The
false-alarm probability (FAP) of this peak is less than 0.001
according to the formulation of \citet{HorneBaliunas1986}.  While this
FAP calculation is strictly applicable only to evenly-spaced data, a
Monte Carlo bootstrapping method \citep[e.g.,][]{Stassun1999} confirms
that this period is statistically significant at better than 99.9\%
confidence.  The period uncertainty reported above, which follows from
the formulation of \citet{Kovacs1981}, is probably underestimated as
it assumes that the underlying signal is well described by a single
sinusoid.

\begin{figure}
\begin{center}
\includegraphics[scale=1.15]{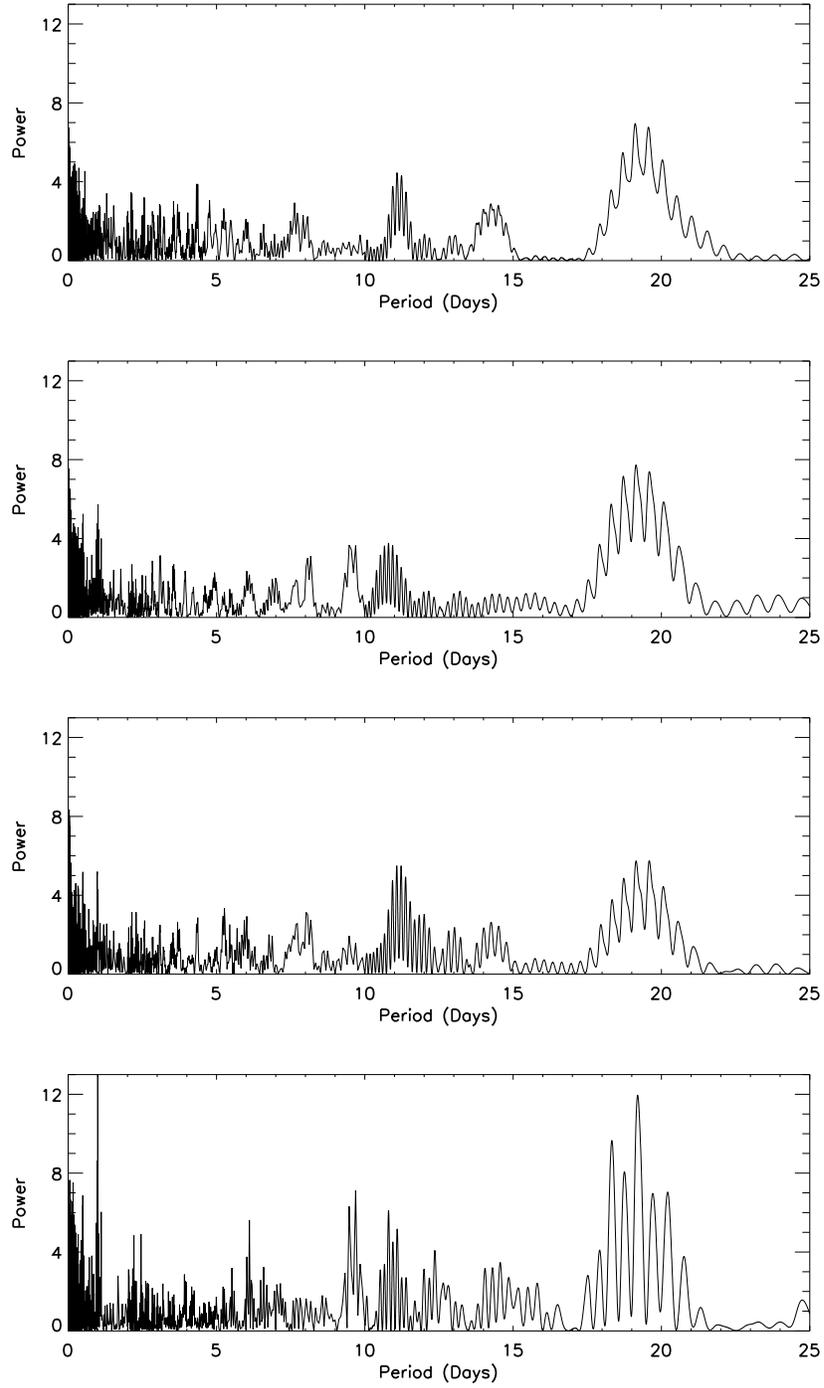}
  \caption{The Lomb-Scargle periodograms for the $B$, $V$, $R$, and $I$-band
    data. All show significant power near the binary orbital
    period.\label{figure:periodogram-bvr}}
\end{center}
\end{figure}

Though it is slightly off the main peak of the power spectrum, there
is significant power at the binary period of 19.131 days.
Periodograms of the $B$, $V$, and $R$ data are similar (Figure
\ref{figure:periodogram-bvr}), showing peaks near the binary period,
but with broader peaks and higher false-alarm probabilities, perhaps
due to the more limited time coverage of the data in those bands.

To refine the period and to better estimate its uncertainty we next
performed a phase dispersion minimization (PDM) analysis
\citep{Stellingwerf1978}, which is particularly well-suited to
periodic variability that is highly non-sinusoidal and/or to data with
large intrinsic scatter, both of which apply to the photometry of
UZ~Tau~E. The PDM search of the $I$-band data yields a best period of
$P = 19.17 \pm 0.05$~d, where the uncertainty was determined
empirically from the $1/e$ folding scale of the PDM merit function.
The same analysis on the $V$-band data gives $P = 19.15 \pm 0.04$~d.

\citet{Schwarzenberg-Czerny1989} argues that a related test, the
one-way analysis of variance, is the most powerful statistic of this
kind for detection of periodic signals.  Applying that test to our
data yields $P = 19.16 \pm 0.03$~d for the $I$-band data, and $P =
19.15 \pm 0.05$~d for the $V$-band data.  Following
\citet{Schwarzenberg-Czerny1989}, the period uncertainty was
determined using a ``post-mortem'' analysis that measures the
1-$\sigma$ confidence interval of the primary periodogram peak,
defined by its width at the mean noise power level of the periodogram
in the vicinity of this peak.  As above, a Monte Carlo permutation
analysis of the light curves confirms that this period is
statistically significant at better than 99.9\% confidence.
Combining these estimates, our best-fit photometric period is $P =
19.16 \pm 0.04$~d, consistent with the binary orbital period.

\begin{figure}
\begin{center}
\includegraphics[scale=1.22]{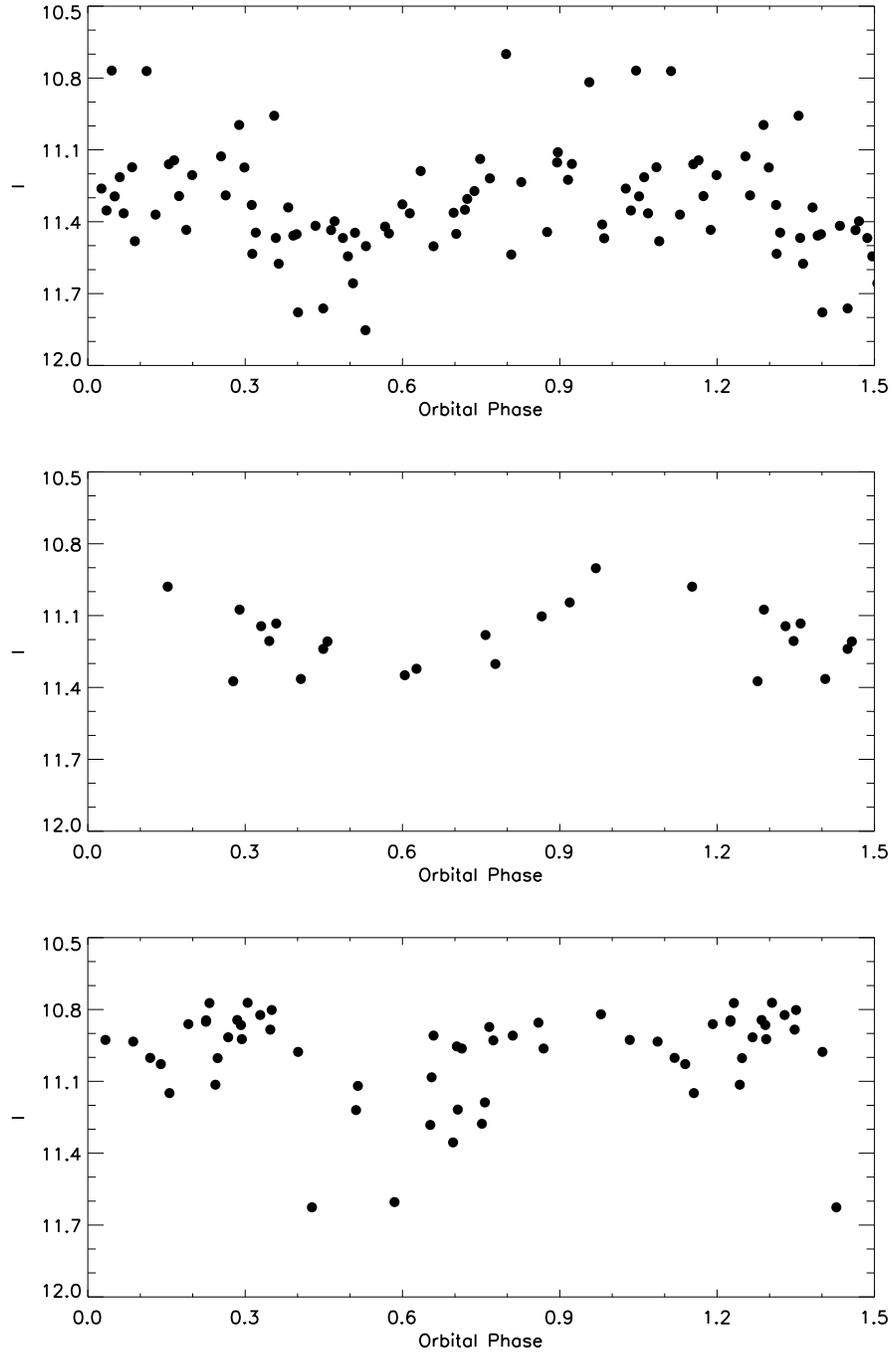}
\caption{The $I$-band magnitude for UZ Tau E folded at the binary orbital
  period of 19.131 days and plotted against the binary orbital phase. 
  Top to bottom, data from 2003--2004, 2004--2005, and 2005--2006.
  \label{figure:i-band}}
\end{center}
\end{figure}

Figure \ref{figure:i-band} shows $I$-band light curves for all three
observing seasons, folded at the binary orbital period of 19.131 days.
As suggested
by the periodogram analysis, all show indications of periodic
behavior, with a broad minimum near orbital phase 0.5.  The data from
the 2004--2005 season show the smoothest variability, but this is also
the season with the smallest number of data points.  Clearly there is
significant random variability as well, with scatter of roughly 0.6
magnitudes at all orbital phases.

\begin{figure}
\plotone{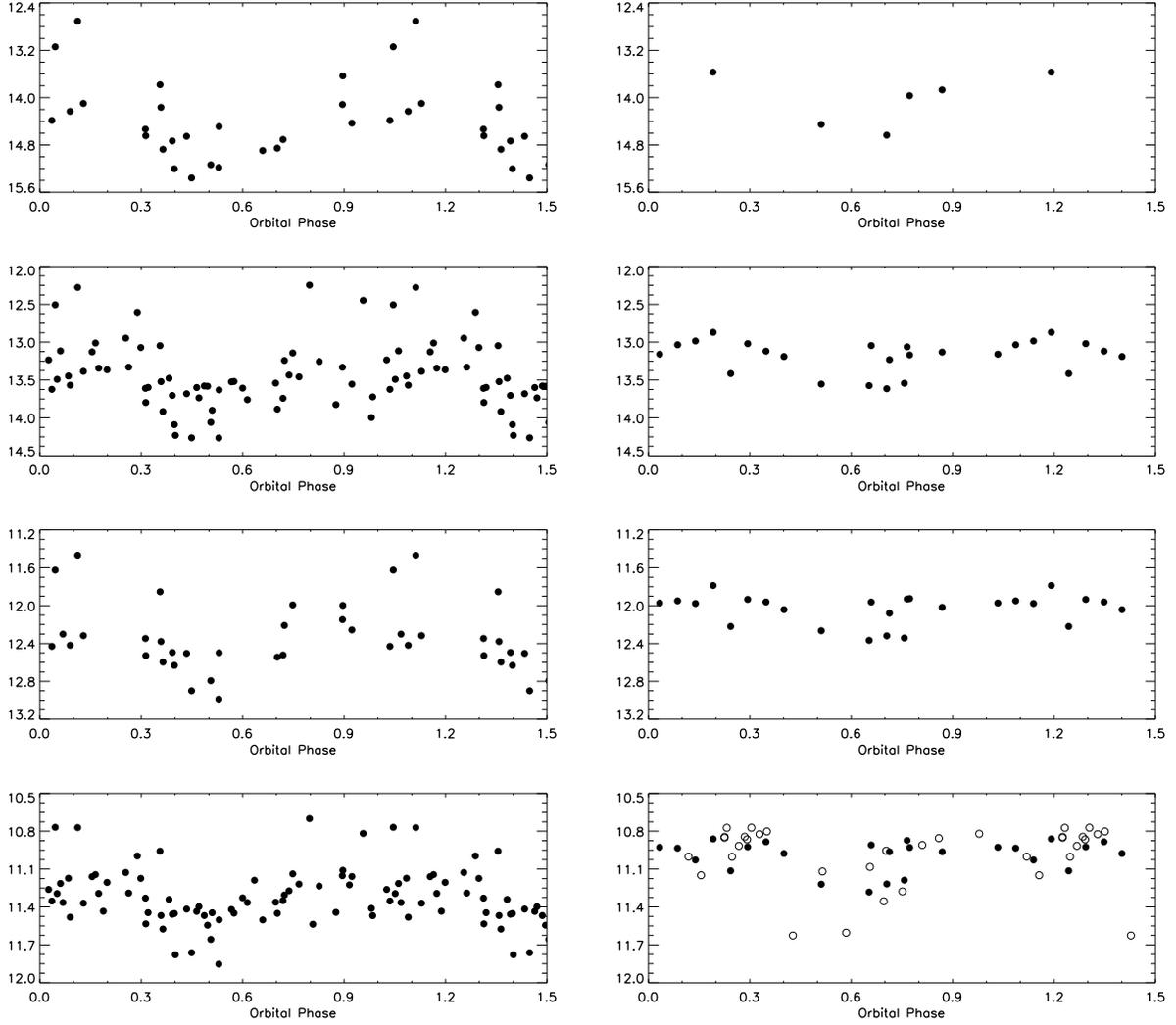}
\caption{The $BVRI$ magnitudes for UZ Tau E folded at the binary orbital
  period and plotted against the binary orbital phase.  
  Left, 2003--2004; right, 2005-2006.  The open
  circles in the lower right plot show the VVO I-band data, which do
  not have corresponding B, V, and R data.
  \label{figure:bvri}}
\end{figure}

Figure \ref{figure:bvri} shows folded $B$, $V$, $R$, and $I$
lightcurves from the 2003--2004 and 2005--2006 seasons.  Broadly
speaking, the $BVR$ data show the same behavior as the $I$-band light
curves, with large-amplitude variability that appears to have both
periodic and random components.  We note that the $R$ band includes
the \ha\ line, which may complicate the interpretation of the light
curve.

\begin{figure}
\begin{center}
\includegraphics[scale=1.13]{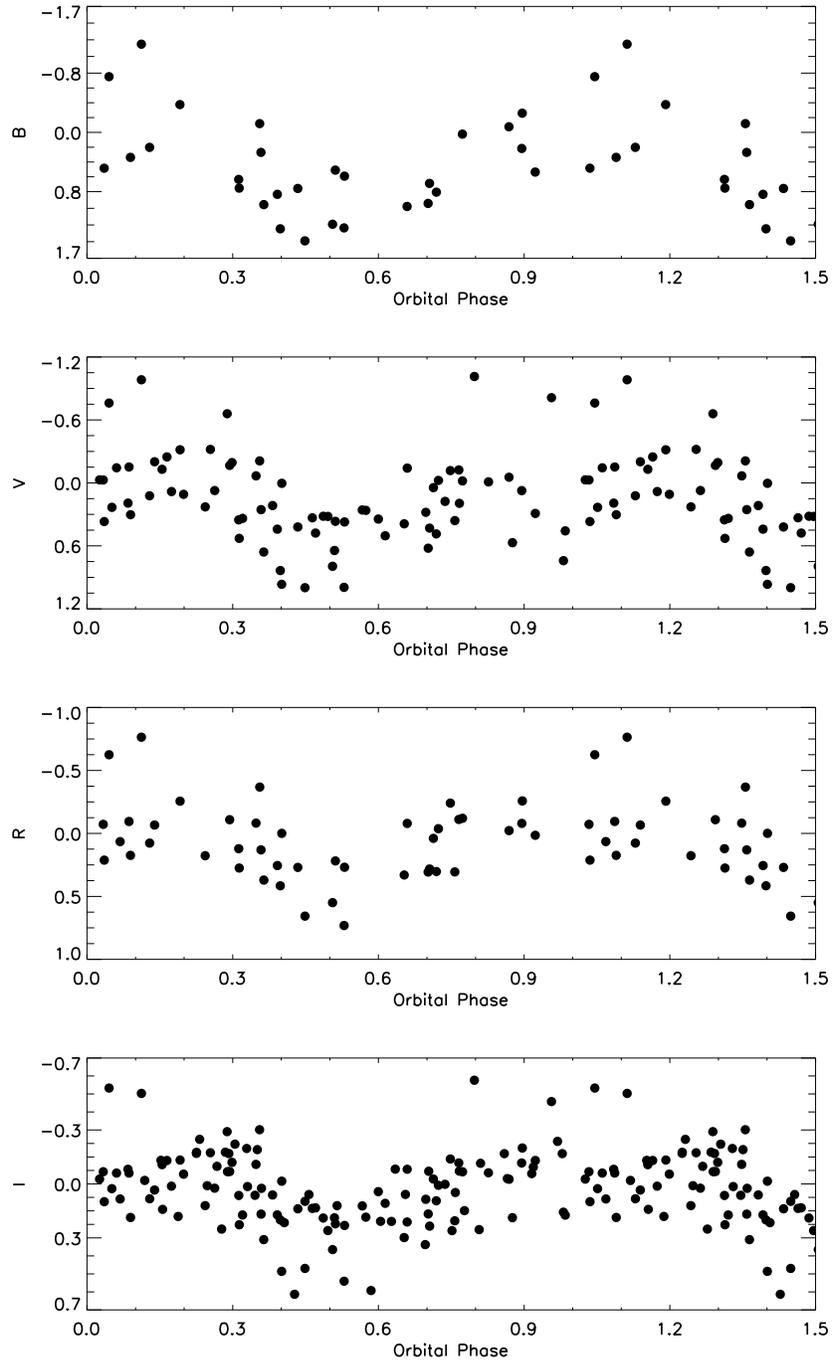}
\caption{The $BVRI$ magnitudes for UZ Tau E folded at the binary orbital
  period and plotted against the binary orbital phase, after removing
  a long-term linear trend from each band.
  \label{figure:bvri-detrended}}
\end{center}
\end{figure}

All four bands show a gradual increase in brightness over the three
years of our observations, with the mean magnitude changing by 0.3
mags at $I$ band from 2003--2004 to 2005--2006.  To separate long-term
variations from the shorter-term variations of interest here, a linear
trend (with a slope of roughly 0.15 mag / yr at $I$ band) has been fit
to each band.  The folded light curves using these de-trended data,
and combining all three observing seasons, are shown in Figure
\ref{figure:bvri-detrended}.

In addition to our photometric data, previous data on UZ Tau have
shown some evidence of both long-term trends and periodic variations.
\citet{Bohlin1923}, in one of the first papers to mention UZ Tau,
reported on a major flare and then a four-magnitude overall decline in
brightness from 1921--1923.  He also noted that there was a
short-period variation with a period of 10--20 days, which encompasses
the period of the variations reported here.  Bohlin's measurements are
for the entire UZ Tau system, but later examination of Bohlin's plates
by \citet{Herbig1977} showed that it was UZ Tau E that brightened
dramatically in 1921.

\begin{figure}
\plotone{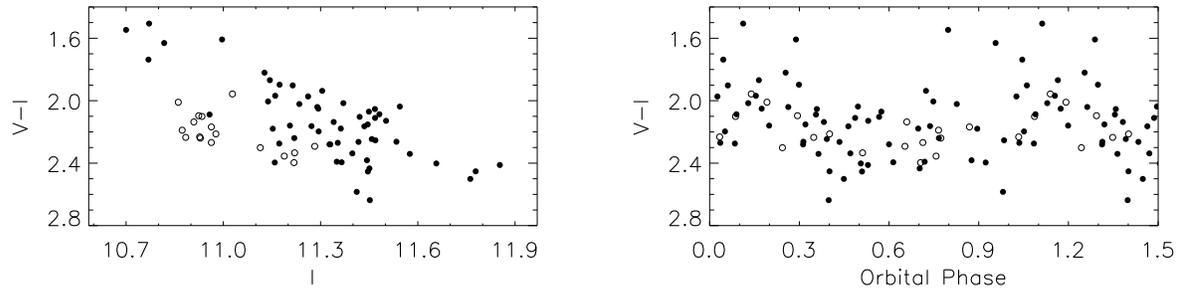}
  \caption{$V-I$ color vs.\ $I$ magnitude and vs. orbital phase for
    2003--2004 (closed circles) and 2005--2006 (open circles).  The
    system is redder when fainter and bluer when brighter, the
    expected behavior either for changes in extinction or for
    brightening due to increased accretion.\label{figure:v-i}}
\end{figure}

Variations in color of the system can also give clues about the cause
of the variability.  Figure \ref{figure:v-i} shows the $V-I$ color as
a function of $I$ magnitude and orbital phase.  The system shows a
behavior commonly seen in T Tauri stars, appearing redder when fainter
and bluer when brighter \citep{Herbst1994}.  This behavior is
consistent either with periodic changes in extinction (causing both
dimming and reddening when the extinction is higher) or in accretion
(adding additional blue light when the accretion rate is higher).

\subsection{Spectroscopy}

If the periodic photometric variations are due to changes in accretion
rate, one might expect accompanying variations in \ha\ emission or
spectral veiling, common tracers of accretion.  \citet{Martin2005}
searched for both of these in UZ Tau E and did not find evidence of
either.  With our new spectra, we can revisit the question of \ha\
variability.

\begin{figure}
\plotone{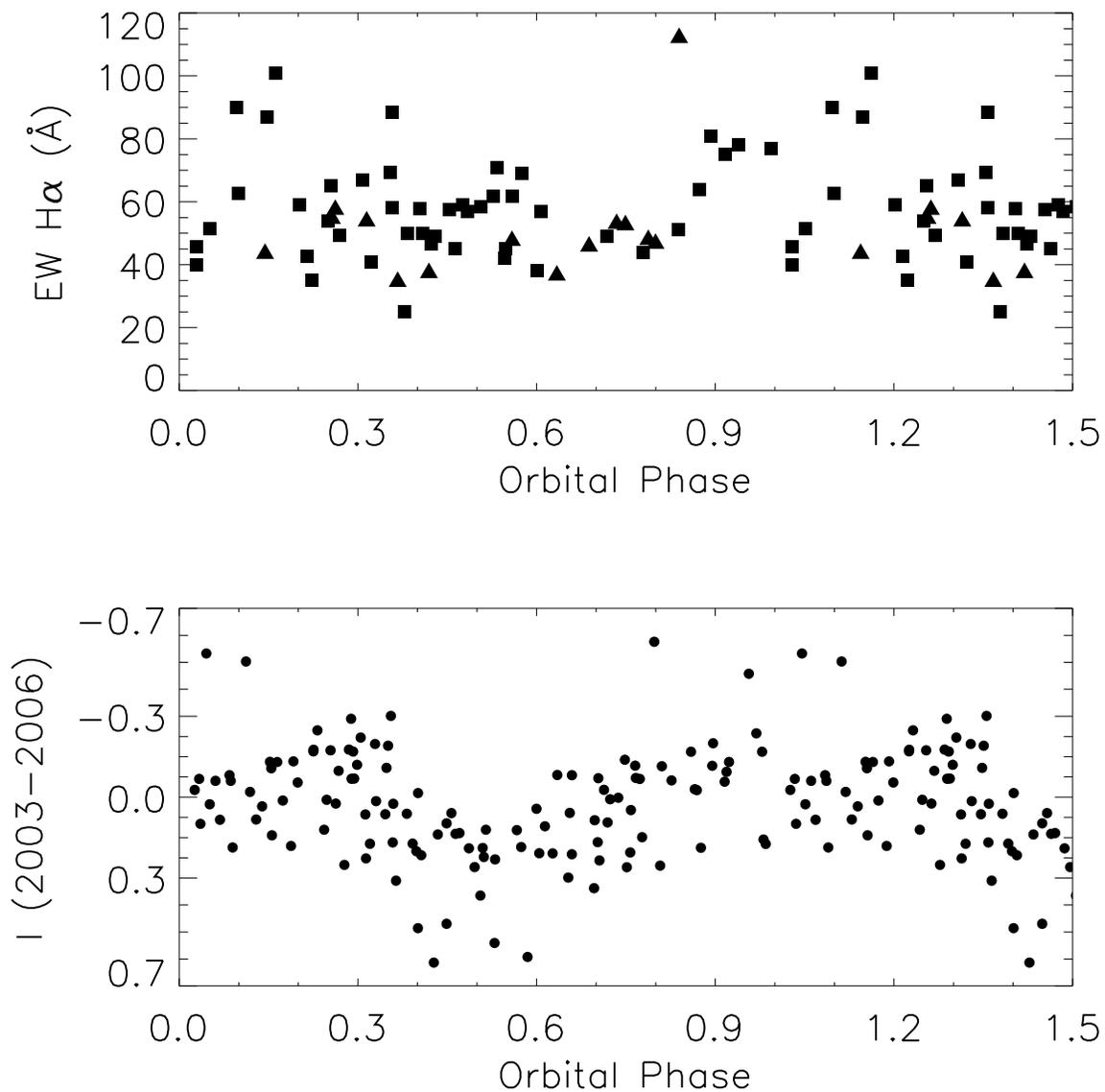}
\caption{Top: Equivalent width of the \ha\ line as a function of
     binary orbital phase.  Squares are our new measurements;
     triangles are from \citet{Martin2005}. Bottom: For comparison,
     the phased $I$-band data.  There is some suggestion of reduced
     \ha\ equivalent width at phases of 0.4--0.8 as seen in the
     photometric data, but the data are too sparse there for there to
     be clear evidence for periodic variability of the \ha\
     emission.\label{figure:halpha}}
\end{figure}

Figure \ref{figure:halpha} shows the equivalent width of the \ha\
emission line as a function of binary orbital phase.  The variations
do not appear to be strongly correlated with orbital phase.  There is
some evidence for lower \ha\ equivalent widths around phase 0.4--0.8,
as seen in the photometric data, but the data are relatively sparse in
that phase range as well.

While a lack of periodic variability would be at odds with the
photometric data, we note that periodicity may not be as obvious in
the spectroscopic data, since the two datasets differ in two important
respects.  First, the \ha\ data have much sparser sampling; they span
a total of eight years (1994--2002), but with only a handful of points
during a given year.  Second, they do not overlap at all with the
photometric data.  Given that the photometric data show both long-term
trends and short-term scatter in addition to the periodic variations,
and that T Tauri stars in general are known to show significant random
variability at \ha, it may be difficult to separate random and
periodic variations (if any) without a dedicated monitoring campaign,
preferably one that includes simultaneous photometric and
spectroscopic measurements.  We conclude that while periodic
spectroscopic variations similar to those seen in the photometry are
not definitively present in these spectroscopic data, neither are they
ruled out.

\section{Discussion}

The photometric data (and possibly the spectroscopic data) show
periodic variability at the binary orbital period, suggesting that
there is a link between the variability and interactions of the binary
with its circumstellar and/or circumbinary material.  In this section,
we first argue that the periodic variations are unlikely to be due to
stellar rotation, and then we examine how well the observed behavior
matches what is expected from the pulsed accretion model of
\citet{AL96}.  Finally, we examine the available data for other
spectroscopic binaries to assess whether or not there is evidence for
periodic accretion as a general phenomenon.

\subsection{Could the variations be due to rotation?}

Periodic variability is not uncommon in photometric studies of PMS
stars. Indeed, dedicated monitoring surveys of rich star-forming
regions \citep[e.g.,][]{MandelHerbst1991,AttridgeHerbst1992,
  ChoiHerbst1996,Stassun1999,Rebull2001,Herbst2002} have now
discovered hundreds of PMS stars exhibiting periodic variability, the
result of surface-brightness inhomogeneities (i.e.\ starspots) that
rotate in and out of view with the stellar rotation period. The
periodic variability observed in UZ~Tau~E is very unlikely to be the
result of such rotationally modulated spot signals, for several
reasons. First, the rotation periods of low-mass PMS stars are nearly
always shorter than about 12 days, while the period of the variations
reported here is 19 days. Among 150 low-mass PMS stars in the Orion
Nebula Cluster, only two stars $(\sim 1\%)$ have $P_{\rm rot} > 15$~d
\citep{Herbst2005}. Second, rotationally modulated spot signals are
typically sinusoidal, and stable over many cycles or, in some cases,
many years. In contrast, the periodic signal we have found in UZ~Tau~E
is decidedly non-sinusoidal, with considerable scatter; the ``bright"
state has a duty cycle of $\sim 60\%$.  Thus, it is either
intrinsically non-sinusoidal, or shows substantial phase shifting from
one cycle to the next; neither of these is consistent with
rotationally-modulated variability.

The rotation period distributions discussed above are presumably
dominated by single stars or member of wide binaries, while tidal
interactions between the stars in a close binary system can
synchronize the orbital and rotational periods.  However, in the case
of eccentric systems like UZ Tau E, pseudo-synchronization (in which
the stellar angular velocity is synchronized with the orbital angular
velocity at periastron) occurs instead, since the tidal interactions
are strongest around periastron \citep{Hut1981}.  The predicted
pseudo-synchronous rotation period for UZ Tau E, using the weak
friction formulation of \citet{Hut1981} and the orbital parameters in
Table \ref{table:orbit}, is $11.4 \pm 1.2$ d, inconsistent with the
observed variability period.

We can also estimate the rotation period of the UZ Tau E primary
directly if three quantities are known: the inclination $i_{\rm rot}$
of the star's rotation axis, the star's projected rotational velocity
$v \sin i_{\rm rot}$, and the stellar radius.  Of these, the
inclination is typically impossible to measure, except under special
circumstances.

As noted in Section \ref{section:orbit}, the dynamical mass
measurement of UZ Tau E allows us to determine the binary orbital
inclination $i_{\rm orbit}$.  Based on studies of other binary
systems, it is reasonable to assume that this inclination is the same
as that of the stellar rotation axis, $i_{\rm rot}$.  The most
detailed study comparing the orientations of these axes in binary
systems is that of \citet{Hale1994}. Considering spectral types of
F5--K5, he finds that binaries with separations less than 30--40 AU
tend to exhibit co-planarity between rotational equators and orbital
planes, while wider binaries have random orientations.  Using a
similar method, \citet{Weis1974} found a tendency for the stellar
rotational equators to align with the binary orbit among primaries in
F star binaries.  Interestingly, \citet{Weis1974} did not find a
tendency toward co-planarity between rotational and orbital planes
among A stars, suggesting that caution is necessary when comparing
stars of different masses.  Similarly, \citet{Guthrie1985} found no
correlation between orbital inclination and $v \sin i$ among 23 A2--A9
binaries with semi-major axes of 10--70 AU\null.  The low mass and
short period of UZ Tau E suggest, however, that the conclusions of
\citet{Hale1994} are most applicable here.

\citet{Prato2002} find $L = 0.63^{+0.19}_{-0.17}$ $L_\odot$ and
$T_{\rm eff} = 3700 \pm 150$ K for the primary in UZ Tau E\null.
Combining these values yields $R = 1.9 \pm 0.2~R_\odot$.
\citet{Hartmann1989} find $v \sin i = 15.9 \pm 4.0$ \kms\ for UZ Tau E
using optical spectra, consistent with the value $v \sin i = 16 \pm 2$
\kms, which we measure from our new spectra and adopt here.  Since
absorption lines of the secondary of UZ Tau E have only been seen in
near-infrared spectra and are not evident in any of our optical
spectra, we take this to be the projected rotation velocity of the
primary.  Combining these measurements with $\sin i_{\rm orbit} = 0.81
\pm 0.05$ (Section \ref{section:orbit}), and assuming $i_{\rm orbit} =
i_{\rm rot}$, we find $P_{\rm rot} = 4.9 \pm 0.8$ d.  If $i_{\rm
  orbit} \ne i_{\rm rot}$, we find $P_{\rm rot} \le 6 \pm 1$ d since
$\sin i \le 1$.  Thus, uncertainty on the inclination cannot reconcile
the photometric period with the inferred rotation period.

The most uncertain remaining quantity is $v \sin i$, but since
\citet{Hartmann1989} measured $v \sin i$ from 11 different spectra of
UZ Tau E, with self-consistent results from two different parts of the
spectrum (including spectra near $\lambda = 5200$ \AA) and consistency
with our new $v \sin i$ measurement, it is unlikely that line
broadening from photospheric lines of the faint, red secondary could
lead to an overestimate of $v \sin i$ by a factor of four.  Similarly,
given the uncertainties on $L$ and $T_{\rm eff}$, it is difficult to
see how the radius could be underestimated by a factor of four.  Thus,
we conclude that the observed periodic variations are unlikely to be
due to stellar rotation.

\subsection{Evidence for pulsed accretion}

We have shown above that UZ Tau E exhibits periodic photometric
variations that have the same period as the binary orbit, and that
these variations are unlikely to be caused by stellar rotation.  Here,
we examine the predictions made by the pulsed accretion model of AL96
and compare them to our observations.

\subsubsection{What are the predictions?}

Broadly speaking, \citet{AL96} predict that a binary with an eccentric
orbit and a circumbinary disk will have an accretion flow from the
circumbinary disk, and thus onto the circumstellar disks or stellar
surfaces, that varies periodically at the binary orbital period.

The exact behavior of the accretion rate with orbital phase depends on
the binary orbital parameters.  \citetalias{AL96} show the results of
two simulations, one for mass ratio $M_2/M_1 = 0.43$ and eccentricity
$e = 0.1$, and another for $M_2/M_1 = 0.79$ and $e = 0.5$.  The former
shows accretion that varies relatively smoothly over the orbital
period, while the latter is strongly peaked at periastron.  As noted
by \citetalias{AL96}, the exact timing of the accretion variability
depends on the orbital parameters, most strongly on $e$.  Some
previous observational studies of T Tauri spectroscopic binaries have
focused specifically on looking for enhanced accretion near
periastron; however, we note here that the actual prediction of the
model is more general than that, and that the peak accretion rate need
not come near periastron.

\subsubsection{How well do the data match the predictions?}
\label{section:pulsed}

First, we note that our observations match the general predictions of
\citetalias{AL96} quite well, in that there are indeed periodic
photometric variations at the binary orbital period, which are readily
interpretable as a variable accretion rate.  The comparison with the
spectroscopic data is more ambiguous; if more intensive monitoring of
the \ha\ line in UZ Tau E were to show that there are no orbit-modulated
\ha\ variations, it would present a problem for the model.

\begin{figure}
\plotone{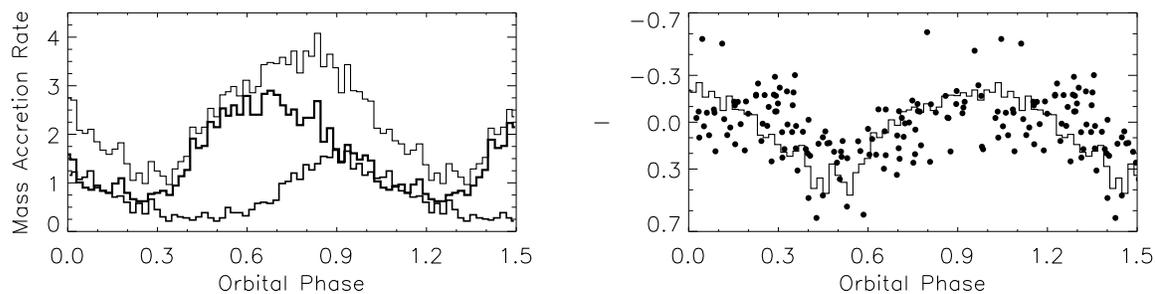}
\caption{Left: The theoretical predictions of \citet{AL96} for the
  dependence of accretion rate on binary orbital phase in a binary
  with mass ratio $M_2/M_1 = 0.43$, $e = 0.1$.  The top curve shows
  the total accretion, while the lower curves show accretion onto the
  secondary (higher dark curve) and primary (lower dark curve).
  Right: The same total accretion curve, but placed onto a logarithmic
  scale and shifted vertically for comparison with the phased $I$-band
  data.  The model here has been given an ad hoc shift of $+$0.2 in
  phase, roughly what is expected given the binary eccentricity (see
  Section \ref{section:pulsed}).\label{figure:al96}}
\end{figure}

For a more specific comparison with our data, Figure \ref{figure:al96}
shows the variations of accretion with orbital phase predicted by
\citetalias{AL96} for a binary with $M_2/M_1 = 0.43$, $e = 0.1$.  UZ
Tau E has a more extreme mass ratio ($M_2/M_1 = 0.30$) and larger
eccentricity ($e = 0.33$) than this, but these parameters are closer
to those of UZ Tau E than those of the other simulation in
\citetalias{AL96}.  \citetalias{AL96} do note that the timing of the
maxima of the accretion depend largely on $e$ rather than $M_2/M_1$.
Since $e$ for UZ Tau E is intermediate between the two models
calculated by AL96, we might then expect the maximum accretion to come
between the phase of $\sim 0.75$ they calculate for the low-$e$ case
and the phase of $\sim 1$ for the high-$e$ case.

For comparison with our data, we have taken the logarithm of the
variations of accretion rate predicted by AL96 to shift them onto a
``magnitude-like'' scale, and added an arbitrary offset and scale
factor to match the mean of the data and amplitude of the variations.
The phase of the minimum predicted by this simulation does not match
our data well; when the model is given a shift of $+$0.2 in orbital
phase, there is better agreement between the model predictions and the
data.  This scaling and shifting to match the data is obviously {\it
  ad hoc}, but it allows us to compare the phase {\em width} of the
observed variations, which appear to match the predictions relatively
well.  In addition, this shifted position of the maximum is indeed
between the two cases calculated by AL96, as expected if eccentricity
is the dominant factor in determining the timing of maximum accretion.

\subsection{Evidence for periodic accretion in other T Tauri binaries}

The discussion and data above show that looking for evidence of
periodic accretion can be complicated, with other sources of
variability perhaps being important and masking the effect in small
datasets, and with the exact behavior expected to be a function of the
specific binary orbital parameters.
That said, is evidence for pulsed accretion seen in other young binary
systems?  In Table \ref{table:ctts-binaries} we present
characteristics of young binaries with periods of less than one year
and evidence of circumbinary material, in order of increasing
eccentricity.  Below, we examine the observational data for some of
these systems, attempting to relate them to what we see in UZ Tau and
exploring similarities and differences.  Unfortunately, the small
number of systems and their somewhat heterogeneous properties means
that it is difficult to generalize, so we offer these comments in the
spirit of attempting to pull together the existing data, rather than
arguing one way or the other for the validity of the \citetalias{AL96}
model for the sample as a whole.

\subsubsection{DQ Tau}

DQ Tau was the first system to be scrutinized for evidence of pulsed
accretion.  \citet{Mathieu1997} showed that the photometric variations
are modulated at the binary orbital period, and \citet{Basri1997}
showed that the \ha\ line and spectral veiling are as well.
Fortuitously, the mass ratio and eccentricity of DQ Tau are quite
similar to those of the high-eccentricity case modeled by
\citetalias{AL96}, allowing for specific comparison with the theory.
The timing and phase width of the photometric and spectroscopic
variations match the predictions well, being sharply peaked near
periastron.  However, the DQ Tau observations did show considerable
orbit-to-orbit variation, with the periastron brightening being seen
roughly 65\% of the time.  This is reminiscent of the large scatter
that we see in the UZ Tau light curves; clearly the periodic accretion
process is not exactly repeatable from orbit to the next, nor is it
the only source of variability.
 
\subsubsection{AK Sco} 

The orbital eccentricity and binary mass ratio of AK Sco are quite
similar to those of DQ Tau, and indeed, simulations by
\citet{GuentherKley2002} for a binary with AK Sco's orbital parameters
predict pulsed accretion.  Thus, it comes as some surprise that the
system does not show periodic photometric variability, despite
extensive monitoring \citep{2003AA...409.1037A}.  The overall
variability is large (up to 1.5 mags in $y$), but apparently random.
There are periodic variations in the Balmer lines, but they are not
sharply peaked around periastron.  Examining Table
\ref{table:ctts-binaries}, we note two properties of AK Sco that are
quite different from those of DQ Tau or UZ Tau.  First, AK Sco is
considerably hotter and more luminous.  Thus, accretion variations of
the same luminosity as those occurring in UZ Tau and DQ Tau would
result in substantially smaller magnitude changes, which could be
swamped by the large random variability.  Second, AK Sco has
considerably lower millimeter flux than either of the other two
systems.  If the systems are fit with similar disk models (in which
the disk is assumed to be optically thin at millimeter wavelengths in
its outer regions), AK Sco's disk mass is an order of magnitude
smaller than that of DQ Tau or UZ Tau E
\citep{JKM1996,JensenMathieu1997,Mathieu1997}.
\citet{2003AA...409.1037A} fit AK Sco with an optically-thick disk
model that has a comparable mass to the disk models fit to DQ Tau and
UZ Tau E.  However, such disk models have not been fit to DQ Tau or UZ
Tau E, and would presumably result in even larger disk masses for
those systems.  In a direct comparison of $\lambda=1.1$ mm flux, DQ
Tau and UZ Tau E are 3 and 5 times brighter than AK Sco at roughly the
same distance, presumably reflecting larger disk masses.  It is
possible that a somewhat lower-mass disk has different dynamics, and
that the accretion flow in the AK Sco system is fundamentally
different than that in the other systems with more massive disks.

\subsubsection{GW Ori} 

This system has a near-circular orbit, and thus would not be expected
to show pulsed accretion under the model set forth by
\citetalias{AL96}.  However, \citet{StempelsGahm2004} quote
Artymowicz, private communication, as saying that pulsed accretion is
possible for systems with circular orbits as well, and indeed
\citet{DAngelo2006} show that this occurs for giant planets embedded
in disks.  Thus, pulsed accretion appears to be possible for at least
some circular-orbit systems and thus may be for GW Ori as well, though
the larger gap cleared by a stellar companion
\citep{1994ApJ...421..651A} will clear some of the disk resonances
that might contribute to disk eccentricity growth in a system with a
planetary-mass companion.


Like AK Sco, GW Ori is very luminous, and shows significant random
variability, though no obvious periodic variability.  It does have a
much more massive disk than AK Sco, however, and indeed than any of
the systems considered here.  Because of its much larger semimajor
axis, and to some extent its circular orbit, GW Ori is much more
likely to have significant {\it circumstellar\/} disks, as the stars
do not approach each other very closely at periastron.  Thus, material
flowing from the circumbinary disk may merge with the circumstellar
disks and then accrete more gradually onto the stars, rather than
falling directly on (or near) the stellar surfaces as is expected to
happen in the shorter-period systems.  If the infalling material does
not shock strongly as it merges with the circumstellar disk, and if
any density enhancements are smoothed out somewhat by the time the
material reaches the stellar surface, then any photometric signature
of the periodic infall would be weakened.  We note that UZ Tau E
likely has circumstellar disks as well \citep{JKM1996}, so a similar
effect could be at work in reducing the amplitude of the periodic
variability relative to the stochastic variability.

\subsubsection{V4046 Sgr}

Like GW Ori, V4046 Sgr has a nearly circular orbit.  However, V4046
Sgr has shown periodic photometric variations at the binary orbital
period \citep{Quast2000,Mekkaden2000}.  These variations persist over
several years and are relatively sinusoidal (Walter, unpublished data,
2003--2005).  Unlike the other binaries discussed here, in this case
stellar rotation is a plausible explanation of the observed
variations.  It is common for stellar rotational periods to become
synchronized with the binary orbital period, particularly for
short-period binaries like V4046 Sgr.  Given the short period
(resulting in stronger tidal interactions and a shorter
synchronization time scale) and the somewhat older age of this system
($\sim 10$ Myr), synchronization is plausible, and indeed is supported
by detailed analysis of the system \citep{StempelsGahm2004}.  However,
rotation does not explain the periodic Balmer line variations
observed, which \citet{StempelsGahm2004} attribute to accumulations of
gas co-rotating with the binary orbit.

\subsubsection{ROXs 42 and ROXs 43B}

These two spectroscopic binaries are both weak-lined T Tauri stars
\citep{BouvierAppenzeller1992, Walter1994}, indicating less-active
accretion than some of the other systems discussed here.  Neither has
been detected at millimeter wavelengths, yielding only an upper limit
on the disk masses \citep{Skinner1991,JMF1996}.  Both systems show
mid-infrared excesses, indicating the presence of circumbinary
material, and a lack of near-infrared excess, which can be modeled as
a cleared central region in the disk \citep{JensenMathieu1997}.  The
fact that both are higher-order multiple systems complicates matters;
ROXs 42 (NTTS 162814$-$2427) is a triple system with a separation of
0\farcs15 \citep{Lee1992, 1993AJ....106.2005G}, while ROXs 43B (NTTS
162819$-$2423S) has a wide companion at 4\farcs8 which is itself a
close binary system \citep{Walter1994,Simon1995}.  Since the evidence
for the presence of a substantial disk rests on the
low-spatial-resolution IRAS detections, it is possible that the excess
is associated with the wider companions rather than arising from
circumbinary disks around the spectroscopic binaries.  In any case,
the lack of millimeter detections indicates that there is less disk
mass in these two systems than in the others discussed here.  Neither
system has been intensively monitored over timespans that would be
necessary to detect periodic photometric variations at the relatively
long orbital periods.  ROXs 42 shows evidence for some semi-regular
variations over roughly 1.5 orbital periods \citep{Zakirov1993}, while
the combined light of the ROX 43 system shows only a 0.1-magnitude
variation, with evidence of a 1.5-day or 3-day periodicity, presumably
attributable to rotation of one or more of the stars
\citep{ShevchenkoHerbst1998}.

\subsubsection{KH 15D} 

The unusual pre--main-sequence system KH 15D (V582 Mon) is a
spectroscopic binary that undergoes deep ($\Delta I \sim 3.5$ mag)
eclipses, thought to arise due to occultation from a circumbinary disk
(\citealt{Hamilton2001,HerbstHamilton2002,Hamilton2005,Winn2006} and
references therein).  While the system has an eccentricity and mass
ratio that would suggest that pulsed accretion might be present, the
photometric variations at the binary orbital period are dominated by
the deep eclipses.  Furthermore, the depth and detailed shape of these
eclipses are evolving with time
\citep{Winn2003,JohnsonWinn2004,Maffei2005,Johnson2005,Winn2006},
making it very difficult to determine whether there might currently be
an additional, smaller-amplitude component with the same period that
is related to accretion rather than occultation.  \citet{Winn2003}
showed that the current deep eclipses did not occur during the first
half of the twentieth century, raising the possibility of searching
for evidence of accretion-related variability at earlier epochs.
Their limit of one mag on the variability during that time does not
preclude accretion-related variations like those seen in UZ Tau
E\null.  The $\sim 0.9$-mag periodic variations seen from the 1960's
through the 1980's \citep{JohnsonWinn2004,Maffei2005,Johnson2005},
however, are relatively smooth and are well-fit by the eclipse model
\citep{Winn2006}, placing a limit on how much any accretion-related
component was contributing to the variability during that time.  Since
the inferred mass ratio and eccentricity for KH 15D are similar to
those of DQ Tau (Table \ref{table:ctts-binaries}), we might expect
accretion-related variability to be strongly peaked around periastron,
which is also when the current deep eclipses occur.  This might help
explain several anomalously bright points seen during eclipses in the
late 1990's that are not well-fit by the model of \citet{Winn2006}.

The precessing circumbinary disk occultation model of \citet{Winn2006}
is quite successful in reproducing the shape and ongoing evolution of
the light curve, and we do not suggest that accretion explains most of
the photometric variations.  We note, however, the possibility that
such an additional component might be sporadically present (with the
same period) and that, if it is, this could complicate the modeling of
the historical evolution of the light curve, especially during
earlier, more-sparsely-sampled epochs.

\section{Conclusions}

We have shown that the pre--main-sequence binary UZ Tau E shows clear
photometric variability at the binary orbital period of 19.13 days.
This variability is consistent with a model in which material in the
circumbinary disk is periodically perturbed by the binary in its
eccentric orbit and falls from the outer disk, across the cleared
central gap and onto the stars or their circumstellar disks.  There is
significant scatter in the light curves, indicating that this ``pulsed
accretion'' may not occur during every binary orbit. \ha\ equivalent
widths show some suggestion of periodic variability, but it is not
definitive.

The apparently intermittent behavior of the accretion, and the
presence of other, random sources of variability, suggest that
searches for this sort of accretion signature require well-sampled
datasets with long time baselines in order to detect any periodic
component.  In particular, simultaneous photometric and spectroscopic
monitoring of UZ Tau E in the future will help determine whether the
\ha\ variations show a periodic component, as the photometric
variations do.  

The good overall agreement between theory and observations suggests
that resonant interactions between stars (and, by extension, planets)
and disks are indeed important in determining disk structure and
dynamics, while the random component of the observed behavior shows
that there is still work to be done in understanding the full
complexity of these interactions.

\acknowledgments

We gratefully acknowledge the support of the National Science
Foundation through grant AST-0307830.  We thank the referee, Steve
Lubow, for useful comments that improved this paper.  We are grateful
to Michael Meyer, David Cohen, and Larry Marschall for useful
discussions; to Marcos Huerta and Pat Hartigan for use of their
spectra of UZ Tau E; to Matthew Richardson for assistance with data
reduction; to Peter Collings for translating early papers on UZ Tau
from German to English; and to Thierry Forveille for use of his ORBIT
code.  MS and FW are grateful for Stony Brook University's partial
support of their participation in the SMARTS consortium.  This
research has made use of the SIMBAD database, operated at CDS,
Strasbourg, France, and of NASA's Astrophysics Data System.


\begin{center}
\begin{deluxetable}{lccccc}
\tablewidth{0pt}
\tablecaption{Observations of UZ Tau E\label{table:observations}
}
\tablehead{
  \colhead{Telescope} & \colhead{Exp.\ time (s)} &
  \colhead{Filter(s)}&\colhead{Season} & \colhead{\# of nights} 
  &  \colhead{Timespan in days}}
\startdata

SMARTS (1.3m)&\phn5  & $BVRI$ & 2003--2004 & 63 &    170\\
              &       &        & 2005--2006 & \phn6 & \phn42\\
              &   30  & $VRI$  & 2005--2006 & \phn9 & \phn33\\
VVO (0.6m)   &   60  &    $I$ & 2004--2005 & 16 & 126\\
             &       &        & 2005--2006 & 20 & 128\\
\enddata
\end{deluxetable}
\end{center}

%
\begin{center}
\begin{deluxetable}{cccc}
\tablewidth{0pt}
\tablecaption{Radial Velocities and H$\alpha$ EW\label{table:spectra}
}
\tablehead{
\colhead{Julian Date}&\colhead{$v_{\rm helio}$ (km s$^{-1}$)}&\colhead{H$\alpha$ EW
  (\AA)\tablenotemark{a}}}
\startdata
 2450416.82& \nodata          & 88.4 \\
 2450783.93& \nodata          & 42 \\
 2450783.94& \nodata          & 45.1 \\
 2450784.96& \nodata          & 38 \\
 2450785.11& \nodata          & 57 \\
 2450835.65& \nodata          & 54: \\
 2451060.99& \nodata          & 45.6 \\
 2451061.00& \nodata          & 39.8 \\
 2451077.14& \nodata          & 63.8 \\
 2451120.92& \nodata          & 101 \\
 2451137.94& \nodata          & 51.3 \\
 2451138.88& \nodata          & 62.6 \\
 2451162.86& \nodata          & 69.3 \\
 2451163.82& \nodata          & 57.8 \\
 2451164.79& \nodata          & 57.5 \\
 2451165.78& \nodata          & 58.3 \\
 2451166.79& \nodata          & 61.7 \\
 2451169.84& \nodata          & 48.9 \\
 2451504.73& \nodata          & 35 \\
 2451507.68& \nodata          & 25: \\
 2451508.67& \nodata          & 49.1 \\
 2451509.71& \nodata          & 57 \\
 2451510.66& \nodata          & 71 \\
 2451517.98& \nodata          & 75 \\
 2451523.68& \nodata          & 42.7 \\
 2451524.74& \nodata          & 49.3 \\
 2451525.76& \nodata          & 40.8 \\
 2451527.68& \nodata          & 46.6 \\
 2451528.68& \nodata          & 58.9 \\
 2451529.66& \nodata          & 61.8 \\
 2451530.60& \nodata          & 69.1 \\
 2452280.60& $-$4.3 $\pm$ 2.1 & 44\\
 2452281.74& $-$3.1 $\pm$ 1.5 & 51\\
 2452282.76& $-$5.8 $\pm$ 1.8 & 81\\
 2452283.66& $-$4.9 $\pm$ 4.6 & 78\\
 2452284.71& \nodata\tablenotemark{b}  & 77\\
 2452286.66& \phn 6.2 $\pm$ 2.6 & 90\\
 2452287.65& 17.1 $\pm$ 1.7 & 87\\
 2452288.68& 27.2 $\pm$ 2.1 & 59\\
 2452289.69& 29.6 $\pm$ 3.0 & 65\\
 2452290.72& 37.8 $\pm$ 7.0 & 67\\
 2452291.67& 28.5 $\pm$ 2.1 & 58\\ 
 2452292.64& 29.1 $\pm$ 3.2 & 50\\
 2452293.67& 22.0 $\pm$ 3.6 & 45\\
 2452579.11& \nodata          & 50 \\
\enddata
\tablenotetext{a}{Positive values denote emission.}
\tablenotetext{b}{The spectrum on this date was too noisy to allow
  measurement of an accurate radial velocity.}
\end{deluxetable}
\end{center}
%

\begin{center}
\begin{deluxetable}{lc}
\tablewidth{0pt}
\tablecaption{Binary orbital parameters for UZ Tau E\label{table:orbit}
}
\tablehead{}
\startdata
Period (days) & $ 19.131 \pm 0.003$\\
$e$             & $ 0.33 \pm 0.04$\\
JD of periastron &  $2451328.3 \pm 0.5$\\
$\omega$         & $239\arcdeg \pm 9\arcdeg$\\
$a \sin i$ (AU)   & $0.124 \pm 0.003$ \\
$\gamma$ (km s$^{-1}$) & $13.9 \pm 0.7$ \\
$K_1$    (km s$^{-1}$) & $17.3 \pm 1.4$ \\
$K_2$    (km s$^{-1}$) & $57.4 \pm 4.7$ \\
$M \sin^3 i$ ($M_\odot$)   &   $0.69 \pm 0.13$ \\
$M_2 / M_1$   &   $0.30 \pm 0.03$ \\
\enddata
\end{deluxetable}
\end{center}
%


\begin{center}
\begin{deluxetable}{lcrccccrcc}
\tablewidth{0pt}
\rotate
\tablecaption{CTTS spectroscopic binaries\label{table:ctts-binaries}
}
\tablehead{
\colhead{Binary}&\colhead{Period}&\colhead{e}&\colhead{$M_2/M_1$}&
\colhead{Spectral}&\colhead{L}&\colhead{Disk Mass\tablenotemark{a}}&\colhead{Photometric}&\colhead{Balmer line}&\colhead{References}\\
\colhead{System}&\colhead{(days)}&\colhead{}&\colhead{}&\colhead{Type}&
\colhead{($L_\odot$)}&\colhead{($M_\odot$)}&\colhead{periodicity?}&\colhead{periodicity?}&\colhead{}}

\startdata
V4046 Sgr&\phn \phn 2.421&$\leq$ 0.01& 0.94&K5&\phn0.82&\phn0.0085&
    Yes ($\Delta$B$\approx$0.1)& Yes & 1, 2, 3
   \\
GW Ori&241.9\phn \phn&0.04&SB1&G0&26\phd\phn\phn&0.3\phn\phn&
    ? ($\Delta$V$\approx$0.7)& ? & 4, 5
   \\
UZ Tau E&\phn 19.131&0.33&0.30&M1&\phn0.91&0.063&Yes
    ($\Delta$I$\approx$0.8)& Maybe & 6, 7, 8
   \\
ROXs 43B  &  \phn 89.1\phn\phn  &	0.41  &  SB1  &  G0   & 0.4    & $<0.00037$   &
     ? ($\Delta$V=0.1)  & ? & 1, 9, 10, 11\\
AK Sco&\phn 13.609&0.47&0.99&F5&\phn8.40&0.002&No
    ($\Delta$y$\approx$1.5)& Yes & 1, 12, 13
   \\
ROXs 42 &	\phn35.95\phn &	0.48  & 0.92  &  K4 &   0.4    & $<0.00025$   &
     ? ($\Delta$V=0.4)   & ?  & 1, 10, 11, 14, 15\\
DQ Tau&\phn 15.804&0.56& 0.97 &K7--M1&\phn0.95&0.020&Yes
($\Delta$V$\approx$0.5)& Yes & 16, 17
   \\
KH 15D&\phn48.38\phn &0.57--0.65&0.83\tablenotemark{b}&K7\tablenotemark{c}&0.4\tablenotemark{c}&\nodata&Eclipse
($\Delta$I$\approx$3.5)& ? & 18, 19, 20\\
\enddata
\tablenotetext{a}{Disk mass estimates were made assuming that the disk is at least
  partially optically thin at mm wavelengths.  An optically-thick
  model for AK Sco  \citep{2003AA...409.1037A} yields a disk mass of 0.02
  $M_\sun$.}
\tablenotetext{b}{Derived from the stellar luminosity ratio that best
  fits the eclipse data \citep{Winn2006}.}
\tablenotetext{c}{Properties of the secondary star, since the primary
  is never visible.}
\tablerefs{    1. \citet{JensenMathieu1997}.
               2. \citet{Quast2000}.
               3. \citet{Mekkaden2000}.
               4. \citet{1991AJ....101.2184M}.
               5. \citet{1995AJ....109.2655M}.
               6. This work. 
               7. \citet{Prato2002}.
               8. \citet{JKM1996}.
               9. \citet{ShevchenkoHerbst1998}.
               10. \citet{MansetBastien2003}.
               11. \citet{BouvierAppenzeller1992}.
               12. \citet{2003AA...409.1037A}.
               13. \citet{2005AJ....129..480M}.
               14. \citet{Lee1992}
               15. \citet{Walter1994}.
               16. \citet{Mathieu1997}.
               17. \citet{Basri1997}.
               18. \citet{Hamilton2001}.
               19. \citet{Hamilton2005}.
               20. \citet{Winn2006}.}
\end{deluxetable}
\end{center}

%

%

%
%
%
%
%

\end{document}